\documentstyle[12pt]{article}

\newcommand{\tseprepno}{Imperial/TP/93-94/39}
\newcommand{\tsehepphno}{hep-ph/9406297}

\def\tsetrue{T} \def\tsefalse{F} 

\let\tsepaper=\tsefalse    
\let\tsenoteon=\tsetrue   
\let\tselse=\tsetrue      
\let\tseletter=\tsefalse  
\let\tsedevon=\tsetrue  

\if\tsetrue\tsedevon \newcommand{\tsedevelop}[1]{{#1}}
\else \newcommand{\tsedevelop}[1]{{}}
\fi

\if\tsetrue\tsepaper 
\else 
\fi
\tsedevelop{\typeout{*** T.S.E. Development mode on ***}}

\if\tsetrue\tsepaper 
\else 
\fi

\if\tsetrue\tsepaper  
\fi

\if\tsetrue\tseletter{
\typeout{          as for Letter paper ---}
}\else
\fi
\if\tsetrue\tselse{\makeatletter

\@addtoreset{equation}{section}
\makeatother
\fi
\if\tsetrue\tselse
\typeout{--- Equations labeled as (section.equation) ---}
\fi

\newcommand{\vol}[1]{{\bf #1}}

\if \tsedevon\tsefalse 
\else 
\typeout{--- Renewing tselabel command ---}
\fi

\newcommand{\tnote}[1]{\if\tsetrue\tsenoteon \footnote{#1} \fi}
\if\tsetrue\tsenoteon{
\typeout{--- Tim Footnotes Included ---}
}\else \typeout{--- Tim Footnotes Excluded ---}
\fi

\newcommand{\tcaption}[2]{\if\tsetrue\tsepaper{
\vspace{5cm} \caption{  }
}\else
{\vspace{#1} \caption{#2} } \fi }

\newcommand{\half}{\frac{1}{2}}
\newcommand{\bea}{\begin{eqnarray}}
\newcommand{\eea}{\end{eqnarray}}
\newcommand{\beq}{\begin{equation}}
\newcommand{\eeq}{\end{equation}}

\typeout{--- Equation break set for wide text ---}
\newcommand{\npagepub} {\if\tsetrue\tsepaper{\newpage }\fi}
\if\tsetrue\tsepaper{
\typeout{--- Page Breaks for Pub. Version ON ---} }\else{
\typeout{--- Page Breaks for Pub. Version OFF ---} }\fi

\newcommand{\quarter}{\frac{1}{4}}

\newcommand{\ket}[1]{| #1 \rangle}
\newcommand{\expect}[1]{\langle #1 \rangle}

\newcommand{\dash}[1]{{#1}^\prime{}}

\newcommand{\veck}{\vec{k}}
\newcommand{\vecp}{\vec{p}}
\newcommand{\vecx}{\vec{x}}

\newcommand{\phibar}{\bar{\phi}}

\newcommand{\vecxdash}{\dash{\vecx}}

\newcommand{\phiv}{\phi_{v}}

\newcommand{\tin}{{t_{in}}}

\newcommand\ba{\begin{array}}
\newcommand\ea{\end{array}}
\newcommand\ben{\begin{equation}}
\newcommand\een{\end{equation}}

\def\fun#1#2{\lower3.6pt\vbox{\baselineskip0pt\lineskip.9pt
\ialign{$\mathsurround=0pt#1\hfill##\hfil$\crcr#2\crcr\sim\crcr}}}

\def\math{\mathsurround 0pt}
\def\oversim#1#2{\lower.5pt\vbox{\baselineskip0pt \lineskip-.5pt

\ialign{$\math#1\hfil##\hfil$\crcr#2\crcr{\scriptstyle\sim}\crcr}}}

\def\({\left(} \def\){\right)}
\def\[{\left[} \def\]{\right]}

\def\half{{\mathchoice{{\textstyle{1\over 2}}}{1\over 2}{1\over 2}{1
\over 2}}}
\def\unit#1{\ifinner \;
            \else \quad \fi
            {\rm #1}}

\setbox1=\hbox{$\mathchar'173$}
\def\dbar{\raise0.3em\copy1\mskip-12mud}
\def\deltabar{\raise0.3em\copy1\mskip-12mu\delta}

\begin{document}

\typeout{--- Title page start ---}

\if\tsefalse\tsepaper \thispagestyle{empty}\fi

\renewcommand{\thefootnote}{\fnsymbol{footnote}}
\begin{tabbing}
\hskip 11.5 cm \= \tseprepno \\
\> \tsehepphno \\
\end{tabbing}
\vskip 1cm

\begin{center}
{\Large\bf Fluctuations at Phase Transitions\footnote{Invited talk
to Nato Advanced Research Workshop on Electroweak Physics and the Early
Universe, Sintra (Portugal), March 1994 }}
\vskip 1.2cm
{\large\bf R.J. Rivers\footnote{E-mail: R.Rivers@IC.AC.UK}\\
Blackett Laboratory, Imperial College, Prince Consort Road,\\
London SW7 2BZ  U.K.}
\end{center}
\if\tsetrue\tsepaper{\begin{center}
Tel: U.K.-71-589-5111 ext. 6980 \\
Fax: U.K.-71-589-9463 \\
\mbox{  }\\
PACS: 11.10-z
\end{center}
}\fi

\npagepub
\vskip 1cm
\begin{center}
{\large\bf Abstract}
\end{center}

We use the closed time-path formalism to calculate fluctuations
at phase transitions, both in and out of equilibrium.  Specifically,
we consider the creation of vortices by fluctuations, of
relevance to the early universe and to $^{4}He$ superfluidity.
\vskip 1cm

\renewcommand{\thefootnote}{\arabic{footnote}}
\setcounter{footnote}{0}

\npagepub

\typeout{--- Main Text Start ---}

\section{Introduction}

It has become increasingly important to understand the nature of
the field fluctuations that are a consequence of the phase
transitions that occurred in the very early universe.  On the one
hand, fluctuations at the GUT era are assumed to have seeded the
large-scale structure visible today.  On the other hand,
fluctuations at the slightly later electroweak transition are most
likely the basis of baryogenesis.

In this talk I shall consider three aspects of fluctuations.  The
first is combinatorical, establishing a path-integral in terms of
which fluctuations can be evaluated.  The second is concerned
with the initial-value problem, exemplified by both equilibrium and
non-equilibrium processes for a real scalar field.
The third provides a non-trivial
application, the creation by fluctuations of (global) vortices in a
symmetry-broken $U(1)$ theory of a complex scalar field.

This application is highly relevant, on two counts.  Firstly, while
there is no unambiguous mechanism for large-scale structure
formation in the universe, arguably one of the least artificial
(originally proposed by Kibble \cite{TK}) is to attribute it to the
spontaneous creation of cosmic strings (vortices) by fluctuations at
the GUT transitions.  [Primordial magnetic fields are created by
field fluctuations in a related way.  See Enqvist, these proceedings
and elsewhere \cite{KE}].

Secondly, very recent experiments \cite{HLM} show that vortices are
naturally generated in superfluid $^{4}He$ on quenching it through its
critical density.  In the Ginzburg-Landau theory of superfluidity
the vortices of $^{4}He$ are the counterparts of global U(1) cosmic
strings (albeit non-relativistic).  These results have been argued  \cite{Z}
as providing direct support for the creation of cosmic strings in the early
universe.
We have some observations on this.

This highly enjoyable meeting arrived a little too early for some of
the work presented here to be completed and my conclusions are
rather more qualitative than I would have liked.
\footnote{I have taken advantage of the few weeks since the meeting
to develop some of the points a little further.  Conclusions are unaltered.}
A fuller discussion will be
given elsewhere, as indicated in the text.

\section{Combinatorics of Fluctuations.}

To be concrete, we begin with the simplest possible theory, that of
a real scalar field $\phi$, with order parameter $\langle\phi\rangle$.  A
natural
measure of the fluctuations of $\phi (t,\vecx)$ is given by the field
averages
\beq
\phiv(t) = \frac{1}{v}\int_{v} d\vecx\;\phi(t,\vecx)
\eeq
over fixed volumes $v=O(l^3)$.  [The symbol $v$ denotes both the position
of the volume in question and its size].  The system as a
whole is taken to have a large volume $V=O(L^3)\gg v$, in which $V$ is
ultimately taken infinite.  Since, at any time, the field $\phi$ is
correlated over some length $\xi$, it is sufficient to take $l\ge\xi$.

To simplify the notation, assume that the possible configurations
$\Phi_{j}(\vecx)$ of $\phi$ in the volume $V$ are denumerable (e.g.
we impose periodic
boundary conditions).  Particular values of $\phiv$ then single out
particular $\Phi_{j}$.

In describing the quantum theory of $\phi$, it is convenient to adopt the
field representation.  Suppose, at some initial time $t=t_{in}$, the
system, described by the state functional $\ket{\Phi_{k},t_{in}}$, is an
eigenstate
of $\hat{\phi}$, eigenvalue $\Phi_{k}$.  That is,
\beq
\hat{\phi} \ket{\Phi_{k},\tin} =
\Phi_{k}(\vecx) \ket{\Phi_{k},\tin}.
\eeq
At a later time $t > t_{in}$, the state has evolved to
\beq
\ket{\Phi_{k},t}=\sum_j c_{kj}\ket{\Phi_j ,t}
\eeq
a superposition of field eigenstates.  The probability that the
measurement of $\hat{\phi}$ in $\ket{\Phi_k ,t}$ gives $\Phi_j$ is
$|c_{kj}|^2$, most conveniently written in terms of path integrals as
\beq
|c_{kj}|^2 = \int_{\phi_1 (t_{in},\vecx) = \Phi_k (\vecx)}
^{\phi_1 (t,\vecx) = \Phi_j (\vecx)} D\phi_1 \exp \{i S[\phi_1]\}
\int_{\phi_2 (t_{in},\vecx) = \Phi_k (\vecx)}
^{\phi_2 (t,\vecx) = \Phi_j (\vecx)} D\Phi_2 \exp \{-iS[\phi_2]\}.
\eeq
The first term in (2.4) is $c_{kj}$, the second $c_{kj}^{*}$, where
$S[\phi] = \int d^4 x {\cal L}(\phi)$
denotes the classical action that determines the evolution of the
system.

The probability $p(\phiv (t) = \phibar)$ that the coarse-grained field $\phiv
(t)$ takes
some chosen value $\phibar$ is obtained by checking which $\Phi_j (\vecx)$ have
their
average as $\phibar$, and multiplying by the probability for finding such
a $\Phi_j$.  That is,
\bea
\lefteqn{p(\phiv (t) - \phibar) = \sum_{j} |c_{kj}|^{2}\delta \left(\frac{1}{v}
\int_{v} d^{3} \vecxdash \Phi_{j}(\vecxdash) -\phibar \right)}
\\
&=& \sum_{j}\int_{\phi_{1} (\tin) =\phi_{2} (\tin)  =\Phi_k  }^{
\phi_{1} (t) =\phi_{2} (t) =\Phi_{j}}
D\phi_1  D\phi_2\;
\delta \left(\frac{1}{v} \int_{v} d^{3} \vecxdash \phi_1(t,\vecxdash) -
\phibar \right)
\exp \{i S[\phi_1] - iS[\phi_2] \}\nonumber
\\
&=& \int_{\phi_{1} (\tin) =\phi_{2} (\tin)  =\Phi_k  }^{
\phi_{1} (t_{fin}) =\phi_{2} (t_{fin})  }
D\phi_1  D\phi_2\;
\delta \left(\frac{1}{v} \int_{v} d^{3} \vecxdash \phi_1(t,\vecxdash) -
\phibar \right)
\exp \{i S[\phi_1] - iS[\phi_2] \}.
\eea
It might seem that the upper bound of the field
integrals in (2.6) should be $\phi_{1} (t,\vecx) = \phi_{2} (t,\vecx)$,
but we can extend the
time contour to any final time value $t_{fin} > t$ without changing the
integral.  We shall take $t_{fin}$ infinite.           .

The expression (2.6) is written more simply as the closed-time path
integral of Schwinger and Keldysh
\bea
p(\phiv (t) - \phibar) &=&
\int_{C} D\phi\; \delta \left(\frac{1}{v} \int_{v} d^{3} \vecxdash
\phi_1(t,\vecxdash) -
\phibar \right)
\exp \{ i S_{ctp} \}
\eea
where $C= C_1\oplus C_2$ is the closed-time path (see below) in which $C_1$
runs from
$t_{in}$ to $t_{fin}$ and $C_2$ runs backward from $t_{fin}$ to $t_{in}$,
infinitesimally
beneath $C_{1}$.
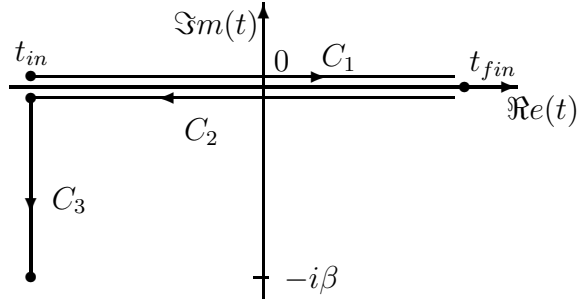
\begin{center}
\setlength{\unitlength}{0.4pt}%
\begin{picture}(500,280)(35,480)
\put( 70,565){\makebox(0,0)[lb]{ $C_3$}}
\put(195,630){\makebox(0,0)[lb]{ $C_2$}}
\put(185,730){\makebox(0,0)[lb]{ $\Im m(t)$}}
\put(325,700){\makebox(0,0)[lb]{ $C_1$}}
\put(280,695){\makebox(0,0)[lb]{ $0$}}
\put(500,650){\makebox(0,0)[lb]{ $\Re e (t)$}}
\put(465,695){\makebox(0,0)[lb]{ $t_{fin}$}}
\put( 35,705){\makebox(0,0)[lb]{ $t_{in}$}}
\put(290,490){\makebox(0,0)[lb]{ $-i\beta$}}
\thicklines
\put( 60,690){\circle*{10}}
\put( 60,670){\circle*{10}}
\put( 60,500){\circle*{10}}
\put(470,680){\circle*{10}}
\put( 40,680){\vector( 1, 0){480}}
\put( 60,690){\vector( 1, 0){280}}
\put(340,690){\line( 1, 0){120}}
\put(460,670){\vector(-1, 0){280}}
\put(180,670){\line(-1, 0){120}}
\put( 60,670){\vector( 0,-1){110}}
\put( 60,560){\line( 0,-1){ 60}}
\put(270,500){\line( 1, 0){ 15}}
\put(280,480){\vector( 0, 1){280}}
\end{picture}
\end{center}
The relevant action  $S_{ctp}[\phi]$, in the presence of
a source $j(t,\vecx)$, is described in terms of ${\cal L}(\phi)$ by
\bea
S_{ctp}[\phi,j] &=& \int_c dt d\vecx ({\cal L}(\phi) + j(t)\phi(t))
\\
&=& \int_{-\infty}^{\infty} dt d\vecx \left( {\cal L}(\phi_1) - {\cal
L}(\phi_2)
 + j_1(t)\phi_1(t) +  j_2(t)\phi_2(t) \right).
\eea
Spatial labels have been suppressed.  The doublets are defined to be
\beq
\phi_a(t)= \protect\left\{
\begin{array}{lcl}
\phi(t) & t \in C_{1} & \mbox{ if $a=1$.}  \\
\phi(t- i \epsilon) & t-i\epsilon \in C_{2} & \mbox{ if
$a=2$.} \end{array} \right.
\eeq
and
\beq
j_a(t)= \protect\left\{
\begin{array}{lcl}
j(t) & t \in C_{1} & \mbox{ if $a=1$.}  \\
- j(t- i \epsilon) & t-i\epsilon \in C_{2} & \mbox{ if
$a=2$.} \end{array} \right.
\eeq
Note that, notionally, we have the second leg $C_{2}$  of our curve
running back $\epsilon$ below the first.

This takes account of the quantum uncertainty.  However, in general,
the initial field  $\Phi(t_{in},\vecx)$ is only specified statistically.  Let
$P_{in}[\Phi_{k}]$   be the probability that, at time $t=t_{in}$, the field
takes
configuration $\Phi_{k} (\vecx)$.  Then
\bea
\lefteqn{p(\phiv (t) - \phibar) = \sum_{k} P_{in}[\Phi_{k}]
\int_{\phi_{1} = \phi_{2} = \Phi_{k}} D\phi_{1} D\phi_{2}\; \delta
\left(\frac{1}{v} \int_{v} d^{3}
\vecxdash \phi_1(t,\vecxdash) -\phibar \right)
\exp \{ i S_{ctp} \}}\nonumber
\\
&=& \int D\Phi P_{in}[\Phi]
\int_{\phi_{1} = \phi_{2} = \Phi} D\phi_{1} D\phi_{2}\; \delta
\left(\frac{1}{v} \int_{v} d^{3}
\vecxdash \phi_1(t,\vecxdash) -\phibar \right)
\exp \{ i S_{ctp} \}.
\eea
where we have now relaxed the (artificial) denumerability of the initial state.
The generality of (2.12) is usually too much for us.  To be
tractable, simple initial conditions are required.  The most
convenient assumption is that $P_{in}[\Phi]$ is Boltzmann distributed as
\beq
P_{in}[\Phi] = N \exp\{-\beta_{in} H_{in}[\Phi]\}
\eeq
at some initial temperature $T_{in} = \beta_{in}^{-1}$.  This permits
\cite{SW} us to introduce
an initial Lagrangian density ${\cal L}_{in} (\phi)$ in terms of which $P_{in}$
can be
expressed as the Euclidean-time path integral (the diagonal
density-matrix element)
\beq
P_{in}[\Phi] = \int_{\phi_{3} (t_{in},\vecx) = \Phi (\vecx)}
D\phi_{3} \exp\{-S_{in} [\phi_{3}]\}
\eeq
where $S_{in} [\phi] = \int dx {\cal L}_{in} (\phi (x))$ is
independent of $T_{in}$.
The time contour for the integral (2.14) is $C_{3}$ (see above), running from
time $t_{in}$
to $t_{in} - i\beta_{in}$, over which $\phi_{3}$ is periodic.  We stress that
${\cal L}_{in} (\phi)$ exists only to parametrise the initial conditions
and, in principle, has nothing to
do with the $\cal L (\phi)$ of later times.  However, in practice
we have in mind a situation in which ${\cal L}_{in}$ and $\cal L$ of the
closed-time path have the same form, but in which the parameters have
a time-dependence \cite{La}.  The other obvious initial condition, in which
the field $\phi$  is localised at a particular value, in effect, can
also be taken as a particular choice of Boltzmann
distribution, as we shall see.  [In principle we can invert (2.14)
to deduce $S_{in}$ from $P_{in}$, whether $P_{in}$ describes a
Boltzmann distribution or not.  In practice, it is all but
impossible.]

With this proviso, $p$  of (2.12) can be written as
\bea
p(\phiv (t) - \phibar) &=&
\int_{C} D\phi_{1} D\phi_{2} D\phi_{3} \delta \left(\frac{1}{v} \int_{v} d^{3}
\vecxdash \phi_1(t,\vecxdash) - \phibar \right)
\exp \{ i S_{C} \}
\eea
where $C$ is the contour $C_{1}\oplus C_{2}\oplus C_{3}$ and $S_{C}
= S_{in}$ on $C_{3}$.

Let $I(\vecx)$ be the window function (indicator function) for the volume
$v$(i.e. $I(\vecx)=1, \vecx\in v ; I(\vecx) = 0, \vecx \not\in v$).
On using an exponential
representation of the $\delta$-function (2.15), rewritten as
\beq
p(\phiv (t) = \phibar) =
\int d\alpha\int_{C} D\phi_{1} D\phi_{2} D\phi_{3}
\exp\{iS_{C} [\phi] -\int_{v} d\vecx\alpha (\phibar
-\phi_{1}(t,\vecx))\}
\eeq
can be further recast as
\beq
p(\phiv (t) = \phibar) = \int d\alpha \exp\{-i\alpha v \phibar\}
Z[\alpha I,0,0]
\eeq
$Z[j_1,j_2,j_3]$, defined by
\beq
Z[j_1,j_2,j_3] = \int D\phi_1 D\phi_2 D\phi_3
\exp \{ i S_{C}[\phi_a,{j}_a] \}.
\eeq
is the generating functional for Green functions in the presence of
sources on all three contours (the extension of $S_{ctp}[\phi_{a},j_{a}]$ of
(2.8) to
include $C_{3}$).  In (2.17) $j_{1} =\alpha I(\vecx)$ is a source coupled to
the field on
$C_{1}$.  The boundary conditions are now $\phi_{1} (t_{in}) =
\phi_{3} (t_{in} -i\beta_{in}),\;\phi_{1} (t_{fin}) = \phi_{2} (t_{fin})$

This splitting into triplets is merely a matter of formal
definition.
The reason for doing this is that it manages to encode the initial
conditions into an expression (2.18) that looks familiar, and is
hence amenable to our usual tricks.  For the circumstances that we
shall consider here, a Gaussian approximation to the fluctuations is
a useful first step.  Adopting the notation
\beq
\expect{F[\phi]} = \int_{C} D\phi_{1} D\phi_{2} D\phi_{3}\;
F[\phi_{1}] \exp\{iS_{C} [\phi ,j=0]\}
\eeq
for all $F[\phi]$, it follows from (2.17) that
\beq
p(\phiv (t) = \phibar) = \int d\alpha \exp\{-i\alpha v\phibar\}
\expect{\exp\{i\alpha v\phiv\}}.
\eeq
The Gaussian approximation is obtained by curtailing the cumulant
expansion
\beq
\expect{\exp\{i\alpha v\phiv\}}
=\exp\{i\alpha v\expect{\phiv} - \half\alpha^{2} v^{2}
[\expect{\phiv\phiv} - \expect{\phiv}^{2}] + O(\alpha^{3})\}
\eeq
at $O(\alpha^{2})$. Performing the $\alpha$ integration gives
\beq
p(\phiv (t) = \phibar) = N\exp\{-\phibar^{2}/2\expect{\phiv\phiv}\}
\eeq
where, for simplicity, we have assumed that $\expect{\phiv} = 0$.  Not
surprisingly, the variance in $\phibar$ is given by the coarse-grained
two-point correlation function.

More generally, we might wish to calculate the probability
$p(f_{v}[\phi] = \bar{f})$. where $f_{v}[\phi]$ is some other coarse-grained
functional
of $\phi$.  In
the same approximation
\beq
p(f_{v}[\phi] = \bar{f}) = N\exp\{-\bar{f}^{2}/2\expect{f_{v}f_{v}}\}
\eeq
again assuming $\expect{f_{v}} = 0$ for simplicity.

\section{The Initial Value Problem}

As we have observed, the hard work has all been neatly tucked away
in the boundary conditions.  The familiarity of the path integral
formalism makes it easy to lose the distinction between the contour
$C_{3}$ (the initial condition) and $C_{1} (C_{2})$ that carries the dynamics.
However, for our first case, thermal equilibrium, the distinction
has been intentionally removed.

\subsection{Thermal Equilibrium.}

This is the most extreme case, in which there is no dynamical
evolution of the system.
Although an unlikely occurrence in the very early universe, thermal
equilibrium has been examined in great detail, in particular through
the thermal effective potential $V(\phi)$ \cite{KL}.  It is the behaviour of
$V(\phi)$
that determines the order of the phase transitions in which we are
interested.  Our approach, given in detail elsewhere \cite{HR},
is close to that of Jona-Lasinio \cite{JL},
who defined the effective potential in terms of probabilities for
coarse-grained
fluctuations of the type proposed in the previous section.  With the
probability $p(\phiv (t) =\phibar)$ now time-independent,
the contours $C_{1}$  and $C_{2}$  can
be shrunk to nothing, leaving the familiar Euclidean-time formalism
on contour $C_{3}$.  We stress that, since thermal equilibrium
corresponds to  making the choice $P_{in}[\Phi]$
of (2.14), where $S_{in} [\phi]$  is now the classical action
$S[\phi]$ of the theory, we have no
further freedom in our initial conditions,  To cite one common
misconception, we cannot now use the loop-expanded effective potential
following
from (2.14) as a 'classical' potential in which we subsequently
consider the evolution of field configurations centred on metastable
minima.  Such an action corresponds to imposing two incompatible
boundary conditions simultaneously.  However, something different,
but not wholly dissimilar, may make sense.  [See Weinberg \cite{EW}, these
proceedings].

Let us return to $P[\Phi]$ of (2.14), corresponding to equilibrium at
temperature $T=\beta^{-1}$.  (With no evolution, all suffices have
been dropped).  Because of the reality of the integrand we can
improve upon the Gaussian approximation.  The integrated probability
satisfies the Chebycheff inequality
\bea
p(\phiv\geq\phibar) &=& \int D\Phi \exp\{-\beta H[\Phi]\}\;
\theta\left(\frac{1}{v}\int_{v} d\vecx\;\Phi (\vecx) - \phibar\right)
\\
&\leq& \int D\Phi \exp\{-\beta H[\Phi]\}\;
\exp\{\beta\alpha\left(\frac{1}{v}\int_{v} d\vecx\;\Phi (\vecx) -
\phibar\right)\}
\eea
for all $\alpha\geq 0$.  Define $\omega_{v} (\alpha)$ by
\beq
\exp\{\beta v\omega_{v}(\alpha)\} =
\int D\Phi \exp\{-\beta H[\Phi] + \beta\int j\Phi \}
\eeq
where $j(\vecx) =\alpha I(\vecx)$, as before.  Then
\beq
p(\phiv\leq\phibar) \leq \exp\{\beta v(\omega_{v}(\alpha)
-\alpha\phibar)\}
\eeq
an inequality minimised for $\alpha =\bar{\alpha}$,
\beq
\frac{\partial\omega_{v}(\bar{\alpha})}{\partial\alpha}
= \phibar .
\eeq
That is
\beq
p(\phiv\leq\phibar)\leq \exp\{-\beta vV_{v}(\phibar)\}
\eeq
where $V_{v} (\phi)$, the Legendre transform of $\omega_{v} (\alpha)$, is the
coarse-grained effective potential.

More specifically, when $\beta\ll m$ we identify $H[\Phi]$ with
$S_{3} [\Phi]$, the
3-dimensional action for the light Matsubara mode, obtained by
integrating over all heavy modes (see Kajantie \cite{KK}, these proceedings).
Thus, but for light-mode self-interactions, $V_{v}(\phibar)$ gives the
usual thermal effective potential $V(\phibar)$ (the
convex hull of the loop-expanded potential) in the large-$v$ limit, when the
inequality (3.6)
becomes an equality.  We note that, if we
were considering a gauge theory, rather than just this simple scalar
theory, the interesting features of the effective potential
(e.g.the order of the transition) are largely determined by the
light gauge field modes, and hence are correctly incorporated in
$V_{v}$.

We shall be much simpler, initially restricting ourselves to a
free scalar field, mass $m$, temperature $T\gg m$. If $\tilde{I}(\vec{k})$ is
the Fourier
transform of $I(\vecx)$ (normalised to $\tilde{I}(\vec{0}) = v = O(l^{3})$)
then
\beq
\expect{\phiv\phiv} = \frac{1}{v^{2}}\int \dbar^{3} k
|\tilde{I}(\vec{k})|^{2} G(\vec{k})
\eeq
where
\beq
G(\vec{k})\simeq\frac{T}{\vec{k}^{2} + m^{2}}
\eeq
is the free-field thermal propagator in this regime.

What we see from this is that, as we might have anticipated, the
effect of coarse-graining is, through $|I(\vec{k})|^{2}$, to impose a
cut-off at $|\vec{k}|<l^{-1}$. $\expect{\phiv,\phiv}$ decreases from
$T/m^{2} v$ at large $v$ ($\tilde{I}(\vec{k})\rightarrow\deltabar (\vec{k})$)
to
$T/Am^{2}v = mT/A$ with $A\simeq 10$, when $l = \xi = O(m^{-1})$.
That is, the variance in $\phi$ is
\beq
(\Delta\phi)^{2}\simeq  mT/A
\eeq
for correlation-volume fluctuations.  The value of $A$ depends a
little on the shape of $V$.  What matters is that it reduces
$(\Delta\phi)^{2}$ by an order of magnitude from our first guess \cite{LL}.
This provides a very useful
guide as to when fluctuations are important.  [We note that $V_{v}(\phi)$
gets steeper as $v$ diminishes].

We are interested in phase transitions, and for a real scalar theory
the only possibility is in the breaking of reflection invariance
$\phi\rightarrow\ -\phi$.  Let us now consider the case when the scalar field
theory has
broken symmetry when cold.  We approximate $H[\Phi]$ by the one-loop
form
\beq
H[\Phi] = \int d\vecx\;[\half (\nabla\Phi)^{2} -\half m^{2}(T)\Phi^{2}
+\quarter\lambda\Phi^{4}].
\eeq
The correlation length $\xi = m(T)^{-1}$, where $m(T)$ is the effective mass
\beq
m^{2}(T) = m^{2}\left( 1 - \frac{T^{2}}{T_{c}^{2}}\right)
\eeq
diverges at the critical temperature $T_{c}$, at which the theory undergoes
a second-order transition.

It is not possible to calculate the coarse-grained $V_{v}(\phi)$ of (2.29)
analytically.  In the symmetric phase ($T > T_{c}$) perturbation
theory is appropriate and we can expand about the Gaussian to
include a quartic term $\lambda_{v}\phi^{4}$ in the
coarse-grained potential $V_{v}(\phi)$.  Its strength $\lambda_{v}$
is, not surprisingly, given in terms of the coarse-grained
four-point correlation function
$\expect{\phi_{v}\phi_{v}\phi_{v}\phi_{v}}$.  Very near to the phase
transition the Gaussian approximation breaks down, but even then it
remains a useful qualitative guide.

In the
symmetry-broken phase perturbation theory breaks down, and even the
Gaussian approximation is difficult.  A Gaussian bound can be made
by weakening the inequality  (3.6)
further by restricting the spatial integral
to $\vecx\in v$ in (3.10).  Without worrying about the details
the end result is the highly sensible bound \cite{HR}
\beq
(\Delta\phi)^{2} = \expect{\phiv\phiv} < \sigma^{2} + \frac{1}{\beta
v m^{2}T}
\eeq

The first term on the right hand side of (3.12) describes the
long-range field correlations in the symmetry-broken phase, the
second the fluctuations about the minima at $|\phi| = \sigma$.  For correlation
volumes $v = O(\xi^{3}) = O(m(T)^{-3})$ the free field result suggests that the
inequality is improved by replacing $Tm(T) = (\beta v m^{2}(T))^{-1}$
by $Tm(T)/A$.  When $Tm(T)/A = O(\sigma^{2}(T))$
there is a significant probability that correlation-volumes of
false vacuum can be created.  This is more transparent if written as
the constraint on the effective thermal coupling strength
$\lambda_{3}$, that
\beq
\lambda_{3} =\lambda T/m(T) = O(1)
\eeq
This condition is seen to be just the Ginzburg criterion \cite{G}, the
condition that the effective thermal coupling $\lambda_{3}$
is becoming large, and signalling that the one-loop form (3.10) is beginning to
break down.
[The failure of the Gaussian approximation in the symmetric phase is
now seen to correspond to $|\lambda_{3}| = O(1)$].

If $H[\Phi]$ describes a theory experiencing a first-order transition,
as the EW transition is expected to do, the situation is more
complicated.  There has been an extensive discussion of the relative
role of fluctuations versus quantum tunnelling by Gleiser, Kolb and
others \cite{GK} (although our earlier
caveat about incompatible boundary conditions is valid here).  The
consensus that thermal fluctuations are inadequate to populate the
symmetry-broken vacuum state is largely confirmed by our analysis,
which will be published elsewhere \cite{BR}.

\subsection{Non-Equilibrium Behaviour.}

The picture that we have given above, in which correlation volumes of
false vacuum are produced with significant probability in the Ginzburg
regime demonstrates how large fluctuations can occur.  However, it may be
positively misleading.
The expansion of the very early universe is so rapid that particles
are separated from one another before they have undergone sufficient
collisions to attain it.  To exemplify the nature of out-of
-equilibrium behaviour, and to offer the strongest contrast to
thermal equilibruim, we consider the case of rapid quenching of
the scalar theory from an initially symmetric state at time $t = t_{in}$ to
a symmetry-broken state, described by the classical action
\beq
S[\phi] = \int d^{4}x\;[\half(\partial\phi)^{2} + \half m^{2}\phi^{2}
+ \quarter\lambda\phi^{4}].
\eeq
More details are given by Lee (these proceedings, and work by Lee,
Boyanovsky, de Vega \cite{B}) in which a similar
calculation is performed, and from which we borrow results.  [Although
motivated by
early universe considerations, for simplicity we stay in
flat spacetime. For a more
realistic temporal evolution in the early universe, see de Vega \cite{dV}
(these proceedings).]

At $t = t_{in}$ we would like the initial state to be symmetric,
with $P_{in}$ strongly
peaked about $\phi = 0$. Despite our caveats about an inability to
define temperature in strongly non-equilibrium environments our earlier
observations suggest that this is most simply
achieved by taking a Boltzmann probability distribution $P_{in}[\Phi] =
N\exp\{-\beta_{in}H_{in}[\Phi]\}$.  $H_{in}[\Phi]$  is
derived, as in (2.14), from an action
\beq
S_{in}[\phi] = \int dx\;[\half(\partial\phi)^{2} -\half
m_{in}^{2}\phi^{2}]
\eeq
for some $m_{in}$.  While $S_{in}$ is a phenomenological action
that induces the required initial
peaking, the effect is the same as if the system were in thermal
equilibrium at temperature $T = T_{in} = \beta_{in}^{-1}$ for all
time $t < t_{in}$.  The
dispersion in $\phi$ at this initial time is given by (31) as
$(\Delta\phi)^{2} = O(m_{in}T_{in})$ on
coarse-graining, chosen as we wish.  The parameter $\beta_{in}$
serves solely to fix this.  For $t > t_{in}$ the evolution of the
initial state is taken to be determined by $S[\phi]$ of (3.14 ).
We take $t_{in} = 0$ for convenience.

In effect, we are taking the theory to be determined for all time by
the action
\beq
S_{t}[\phi] = \int d^{4}x\;[\half(\partial\phi)^{2} + \half m^{2}(t)\phi^{2}
+ \quarter\lambda (t)\phi^{4}]
\eeq
where
\beq
m^{2}(t)= \protect\left\{
\begin{array}{lcl}
m_{in}^{2} & t < 0 \\
-m^{2} & t > 0
\end{array} \right.
\eeq
and
\beq
\lambda(t)= \protect\left\{
\begin{array}{lcl}
0 & t < 0 \\
\lambda & t > 0
\end{array} \right.
\eeq
subject to the condition that, for $t < 0$, it was in thermal
equilibrium.  In this extreme case it is apparent that the
equilibrium thermal effective potential $V$ and its coarse-grained
descendents $V_{v}$ have no role to play and it
makes little sense to talk about the order of the transition.

We could have taken $\lambda$ constant for all time, but have not
done so in order to delineate between its role in establishing the initial
condition (for which it is an artefact) and its role in the subsequent
dynamics, in which it is ultimately a constant of the universe.
Unsurprisingly, taking interactions into account properly is hard.
In practice, because of the failure of perturbation theory,
it is difficult to do better than a self-consistent
(Hartree) linearisation of (3.16).  Despite our preamble, for our purposes it
is
sufficient to set $\lambda (t) = 0$ in (3.18) (whence the classical potential
is an upturned parabola for $t > 0$) but only to consider the evolution of the
system to times at which the field configuration would have spread
from $\phi = 0$ to the point of inflexion $\phi_{inf} = O(m/\sqrt{\lambda})$.
Since the
characteristic growth rate is $\phi /m = e^{mt}$, this means times for which
$t= O(m^{-1} ln(1/\lambda))$.

The spinodal dispersion in $\phi$ is now given by the 'free-field'
equal time correlation function (explicitly dependent on $t$)
\bea
\expect{\phiv\phiv} &=& \frac{1}{v^{2}}\int_{\vecx,\vecxdash\in v}
d\vecx\; d\vecxdash\;\expect{\phi(t,\vecx)\phi(t,\vecxdash)}
\\
&=& \frac{1}{v^{2}}\int d\vecx\; d\vecxdash\;I(\vecx)I(\vecxdash)
G(\vecx -\vecxdash;t)
\\
&=& \frac{1}{v^{2}}\int \dbar\veck\;|\tilde{I}(\veck)|^{2} G(\veck;t)
\eea
in terms of the Fourier transform
\beq
G(\veck ;t) = \int d\vecx G(\vecx ;t) e^{-i\veck .\vecx}.
\eeq
At equal-time all correlation functions are the same, once the
boundary condition $G(\vecx ;t_{in}) = G(\vecx ;t_{in} -i\beta_{in})$
has been implemented.

$G(\veck ;t)$ possesses both oscillatory modes and exponentially
growing (fading) modes, as solutions to
\beq
({\partial}_{t}^{2} + \veck^{2} + m^{2}(t)){\cal U}_{k}(t) = 0
\eeq
from which it is constructed.  The dominant contributions to  (3.21)
are the exponential modes, and if we take them only then \cite{B}
\beq
G(\veck ;t)\simeq \frac{\theta (m^{2}
-\veck^{2})}{2\omega_{in}(\veck)}
\left( 1 + \half\left(\frac{m^{2} + m_{in}^{2}}{m^{2} -\veck^{2}}\right)
(cosh(2t\sqrt{m^{2} - \veck^{2}}) - 1)
\right)coth[\half\beta_{in}\omega_{in}(\veck)]
\eeq
where $\omega_{in}(\veck) = \sqrt{\veck^{2} + m_{in}^{2}}$.  If we consider
fluctuations coarse-grained to volumes $v = O(l^{3})$ the
momentum is cut off at $|\veck| = O(l^{-1})$, whereas the $\theta$-function
imposes a
cut-off at $|\veck| = m$. Since we are only interested in $l\geq m^{-1}$ the
$\theta$-function is irrelevant.  On converting to spherical polars in
(3.22) $k^{2} G(\veck :t)$ has, for large values of $mt$, a sharp peak at
$k^{2} = O(m/t)$.
For this to be inside the integral, so that the probability is
significant, requires $l^{2} < t/m$.  Thus the field is correlated on a
scale $\xi = O(\sqrt{t/m})$.  At the relevant time $t = m^{-1} ln(1/\lambda)$
at
which domain formation rapidly slows to a halt, the domain size is then
\beq
\xi = O(m^{-1} (ln(1/\lambda))^{\half})
\eeq
This is in contrast to $\xi = O(m^{-1}/\sqrt{\lambda})$ in the equilibrium case
at the
Ginzberg temperature (3.13).  In practice, unless $\lambda$  is extremely
small the sizes can be comparable when prefactors are taken into
account properly.
Further details are given by Lee.

\section{Vortex Creation from Fluctuations.}

Consider a field theory invariant under a group $G$.   Suppose the effect
of spontaneous symmetry-breaking is to  reduce  $G$ to its subgroup $H$.
The vacuum manifold $M$ can be identified
with the coset space $G/H$.  If $M$ has a non-trivial first homotopy group
$\Pi_{1} (M)$ (i.e.
the embeddings of loops in $M$), then vortices (strings) can
form \cite {TK}.  [If the second homotopy group
$\Pi_{2} (M)$ is non-trivial we can have monopoles, and if the third
homotopy group $\Pi_{3} (M)$ is non-trivial we can have textures].
As we said earlier, the production of vortices in the early universe has the
capacity to seed large-scale structure through their gravitational effects.
In general, GUT vortices are assumed to arise from the breaking of a
local gauge theory.  However, because of their relative simplicity,
a considerable effort has gone into understanding global vortices,
and it is they that we shall mainly consider here.  Given that the
vortices of superfluid $^{4}He$ are global vortices, they are of
more than theoretical interest.

The simplest theory permitting such vortices is the global $U(1)$ theory of
a complex scalar field $\phi$, with Minkowski space-time action
\beq
S[\phi] = \int d^{4}x\;[\half(|\partial\phi|)^{2} + \half m^{2}|\phi|^{2}
- \quarter\lambda|\phi|^{4}].
\eeq
With $m^{2} > 0$ the $U(1)$ symmetry is spontaneously broken, and the
classical vacuum determined by $|\phi|^{2} = \sigma^{2} = m^{2}/\lambda$.
           .
The  vacuum manifold $M = S^{1}$ has homotopy group $\Pi_{1} (S^{1}) =
Z$, and vortices can exist with integer winding number $n$,
Vortices with winding number $|n| >1$ are unstable to decay into
vortices with $|n| = 1$.

Consider an open surface $S$  with oriented boundary $\partial S$.
The line integral
\beq
N_{S} (t) = \frac{-i}{2\pi} \int_{\partial S}
\underline{dl} .\frac{\phi^{\dagger}
\stackrel{\leftrightarrow}{\partial}\phi}
{|\phi|^{2}}
\eeq
measures the winding number of the field configuration on $S$ at time
$t$. (i.e. $2\pi N_{S}$ is the change in phase of the field $\phi$ as it is
taken
once around $\partial S$).

If $N_{S}\neq 0$, continuity of $\phi$  requires that it vanish at some point
or points of $S$.  A vortex of winding number $n$ is a tube of 'false'
vacuum  $(\phi\simeq 0)$ containing a line of zeros of $\phi$  for which
$N_{S}$ of (4.2) has value $n$ for any closed path enclosing this line.  For a
large loop $\partial S$, $N_{S}$ measures the net winding number of the
vortices
that pass through it.  At a distance $r$ from the vortex centre $|\phi|$
approaches its
vacuum value $\sigma$ as $|\phi| = \sigma (1 - O(e^{-mr}))$.  In practice it is
more
convenient to evaluate the related quantity
\bea
\bar{N}_{S} (t) &=& \frac{-i}{2\pi} \int_{\partial S}
\underline{dl} .\frac{\phi^{\dagger}
\stackrel{\leftrightarrow}{\partial}\phi}
{\sigma|\phi|}
\\
&=& \frac{-i}{2\pi\sigma^{2}} \int_{S} \underline{dS}'
.(\underline{\partial}\phi^{\dagger}\wedge\;\underline{\partial}\phi)
\frac{\sigma}{|\phi|}
\\
&=& \frac{2}{2\pi\sigma} \int_{S} \underline{dS}'
.(\underline{\partial}\rho\wedge\underline{\partial}\chi)
\eea
where, in (3.5), we have used the radial/angular decomposition
\beq
\phi = \rho e^{i\chi}
\eeq
of the field.  For a large loop $\partial S$  the difference between $N_{S}$
and
$\bar{N}_{S}$ (not integer) is vanishingly small if no vortices pass close to
$\partial S$,
and $\bar{N}_{S}$ remains a good indicator of vortex production.

On further decomposing the radial field $\rho$ as $\rho = \sigma + h$ for Higgs
field $h$, $S$ of (4.1) becomes
\beq
S[h,\chi] = \int dx\;[\half(\partial h)^{2} + \half\sigma^{2}
(\partial\chi)^{2} - m^{2}h^{2} - \lambda\sigma h^{3} -
\quarter\lambda h^{4}]
\eeq
In the Gaussian approximation the probability that $\bar{N}_{S}(t)$   takes the
value $n$ is, from (2.23)
\beq
p(\bar{N}_{S} (t) = n) = \exp\{-n^{2}/2\expect{\bar{N}_{S}
(t)\bar{N}_{S} (t)}\}
\eeq
On further defining the Goldstone mode $g$ by $g =\sigma\chi$, from
(4.5) it follows that
\beq
\expect{\bar{N}_{S} (t)\bar{N}_{S} (t)} =
(\frac{2}{2\pi\sigma^{2}})^{2}
\int\int_{S} dS' dS''\expect{(\partial h'\wedge\partial g')(\partial
h''\wedge\partial g'')}
\eeq
All fields are at time t.  The primes (doubleprimes) denote fields in the
infinitesimal areas $dS', dS''$ of $S$ respectively.
For economy of notation we have not made ths scalar products explicit.
Without loss of generality we take $S$  in the 1-2
plane, whence
\beq
\expect{\bar{N}_{S} (t)\bar{N}_{S} (t)} =
(\frac{2}{2\pi\sigma^{2}})^{2}\int\int_{i,j = 1.2} dS' dS''
\expect{\partial_{i} h'\partial_{i} h''\partial_{j} g'\partial_{j}
g'' -\partial_{i} h'\partial_{j} h''\partial_{j} g'\partial_{i} g''}
\eeq
It is convenient to refine our notation
further, decomposing space-time as $x = (t,\vecx) =
(t,\vecx_{L},x_{T})$
where $\vecx_{L} = (x_{1},x_{2})$
denotes the co-ordinates of $S$, and $x_{T} = x_{3}$ the transverse
direction to $S$.
Similarly, we separate 4-momentum $p$  as $p = (E,\vecp_{L},p_{T})$.
    .

Let $G_{h}(t,\vecx' -\vecx'') = \expect{h(t,\vecx')h(t,\vecx'')}$,
$G_{g}(t,\vecx' -\vecx'') = \expect{g(t,\vecx')g(t,\vecx'')}$
be the Higgs field and Goldstone mode correlation functions
respectively.  As a first step we ignore
correlations between Higgs and Goldstone fields.  That is, we retain
only the disconnected parts of $\expect{\bar{N}_{S}\bar{N}_{S}}$.  Eqn. (4.10)
then
simplifies to
\beq
\expect{\bar{N}_{S} (t)\bar{N}_{S} (t)} =
(\frac{2}{2\pi\sigma^{2}})^{2}\int\int_{i,j = 1.2} dS' dS''
[\expect{\partial_{i} h'\partial_{i} h''}\expect{\partial_{j} g'\partial_{j}
g''} -\expect{\partial_{i} h'\partial_{j} h''}\expect{\partial_{j}
g'\partial_{i} g''}]
\eeq
which can be written as
\beq
\expect{\bar{N}_{S} (t)\bar{N}_{S} (t)} =
(\frac{2}{2\pi\sigma^{2}})^{2}\int\int\dbar^{3} p'\dbar^{3} p''\;
G_{h}(\vec{p}',t)G_{g}(\vec{p}'',t)
|\tilde{I} (\vec{p}_{L}'' -\vec{p}_{L}')|^{2}
[(\vecp_{L}')^{2}(\vec{p}_{L}'')^{2} - (\vec{p}_{L}' .\vec{p}_{L}'')^{2}]
\eeq
In (4.12) $\tilde{I} (\vec{p}_{L})$ is the Fourier transform of the window
function
$I(\vecx_{L})$ of the surface $S$ (i.e. $I(\vecx_{L}) = 1$ if
$\vecx_{L}\in S$, otherwise zero).  The $G(\vec{p}, t)$ are defined as
in (3.22).

We coarse-grain in the transverse and longitudinal directions by
imposing a cut-off in three-momenta at $|p_{i}|<\Lambda = l^{-1}$, for some
$l$, as
before.  Thus $\bar{N}_{S}$   is now understood as the average value over a
closed set of correlation-volume 'beads' through which $\partial S$    runs
like
a necklace.  However, we leave the notation unchanged.

For large loops $\partial S$, $\tilde{I}
(\vec{q}_{L})\simeq\deltabar (\vec{q}_{L})$, enabling us to write
\beq
\expect{\bar{N}_{S} (t)\bar{N}_{S} (t)} =
(\frac{2}{2\pi\sigma^{2}})^{2}\int\dbar q_{T}\int\dbar^{3} p\;
G_{h}(\vec{p} +\vec{q}_{T},t)G_{g}(\vec{p},t)\int\dbar^{2} q_{L}
|\tilde{I} (\vec{q}_{L})|^{2}
[(\vecp_{L})^{2}(\vec{q}_{L})^{2} - (\vec{p}_{L} .\vec{q}_{L})^{2}]
\eeq
By $\vec{p} +\vec{q}_{T}$ we mean $(\vec{p}_{L}, p_{T} + q_{T})$.
The dependence on the contour $\partial S$
is contained in the final integral
\bea
{\cal J} &=&\int\dbar^{2} q_{L}\;
|\tilde{I} (\vec{q}_{L})|^{2}
[(\vecp_{L})^{2}(\vec{q}_{L})^{2} - (\vec{p}_{L} .\vec{q}_{L})^{2}]
\\
&\simeq& \pi p_{L}^{2}\int^{\Lambda} dq_{L}\; q_{L}^{3}|\tilde{I} (q_{L})|^{2}
\eea
If this is evaluated for a circular loop of
radius $R$, we find
\beq
{\cal J} = p_{L}^{2}\;O(2\pi R/l)
\eeq
as we might have anticipated.  The rms winding number behaves with
path length as $\Delta n \propto {\cal J}^{\half} = O(L^{\half})$
, where $L$  is the number of steps of
length $l$.  This is consistent with the caorse-grained field phases of
different
volumes $v$ being randomly distributed.

The final step is to relate the magnitude of the fluctuations in
winding number $N_{S}$ to the magnitude of the radial (Higgs) field
fluctuations and angular (Goldstone) field fluctuations.  To see
this requires a specific choice of initial conditions, and we repeat
those of the previous section, thermal equilibrium and instantaneous
quenching.

\subsection{Thermal Equilibrium.}

To estimate the thermal fluctuations in $\bar{N}_{S}$ in equilibrium we {\it
a}) neglect
the positivity of $\rho$ and the Jacobean from the non-linear
transformation (4.6) and {\it b}) the non-singlevaluedness of $\chi$.
While valid for small fluctuations around the global minima this can
only be approximate for large fluctuations.  With this proviso,
at temperature $T$ the equilibrium propagators are time-independent,
read off from (3.7) as
\bea
G_{h}(\vec{p}) &=& \frac{T}{\vec{p}^{2} + m_{H}^{2}(T)}
\\
G_{g}(\vec{p}) &=& \frac{T}{\vec{p}^{2}}
\eea
where $m_{H}(T) =\sqrt{2}m(T)$ is the effective Higgs mass at
temperature $T$.  If we take
$\vec{p}_{L}^{2} = 2\vec{p}^{2}/3$ in the integral
then, up to numerical factors, we can approximate (3.10) as
\beq
\expect{\bar{N}_{S}\bar{N}_{S}} \simeq
{\cal J} \left(\frac{2}{2\pi\sigma^{2}}\right)m(T)T
\int_{|\vec{p}|<m(T)} \dbar\vec{p} G_{h}(\vec{p})
\eeq
where we have coarse-grained to the Higgs correlation length $\xi =
m_{H}^{-1}(T)$.
[We have further assumed that the $q_{T}$ integration can be
approximated by setting $q_{T}$ to zero in the integrand.
Qualitatively this is a reasonable simplification].
The integral in (4.19) is essentially the integral (3.21).  The end
result is that, after substitution,
\beq
\expect{\bar{N}_{S}\bar{N}_{S}} =
O\left(\left(\frac{m_{H}(T)T}{\sigma^{2}(T)}\right)^{2}\right)
O(L)
\eeq
where L is the length of the path in units of $\xi$.
Equivalently, on using our previous results for equilibrium
\beq
\expect{\bar{N}_{S}\bar{N}_{S}} =
O\left(\left(\frac{\expect{h_{v}h_{v}}}{\sigma^{2}(T)}\right)^{2}\right)
O(L)
\eeq
One power of $\expect{h_{v}h_{v}}$ comes from $G_{h}$, the other from residual
factors.
 That is, the phase fluctuations are scaled by the Higgs
fluctuations.  In the Ginzburg regime, when correlation-volumes
of the Higgs field can fluctuate to the false vacuum with
significant probability, the fluctuations in field phase on the same
distance scale are of order unity and, from the $O(L)$ term, are
distributed randomly.  [It is unclear whether the inclusion of connected
correlation functions - linking the Higgs and Goldstone fields -
would change the results qualitatively in this regime, and is under
examination.  The inclusion of four-point correlation functions in
the real scalar field calculations had no dramatic effect, although
the circumstances were somewhat different.]  This ability not to just produce
'beads' of false vacuum, but to string them as vortices, is the
Kibble mechanism referred to earlier \cite{TK}, a common starting-point
for numerical calculations \cite{VV}.

A similar analysis can be performed for $^{4}He$, treated as a
quantum-mechanical system of non-relativistic bosonic point-particles
with pair-wise potentials $V(\vecx_{i} -\vecx_{j})$.  In thermal
equilibrium it is well-known \cite{W} that the grand canonical partition
function
for such a system is identical to the partition function
\beq
Z = \int D\Phi\;\exp\{-S[\Phi]\}
\eeq
of a
complex non-relativistic quantum field $\Phi$.  The action in Euclidean
space-time (with
Euclidean time $0\leq\tau\leq\beta$) for particles of mass $m$ and
(effective) chemical potential $\mu$ is
\bea
S[\Phi] &=& \int d^{4}x \Phi^{\dagger} (x)\left(-\frac{1}{2m}\nabla^{2}
-\mu +\frac{\partial}{\partial\tau}\right)\Phi (x)\nonumber
\\
&+& \half\int d^{4}x\int d^{4}x'\;|\Phi (x)|^{2} V(x -x')|\Phi (x')|^{2}
\eea
$V(x) = \delta(t)V(\vecx)$ is the time-instantaneous generalisation of the
two-body
potential introduced earlier.

Equation (4.23) is exact.  If we adopt a local approximation for the two-body
forces
that enables us to replace the final term in (4.23) by $\lambda |\Phi
|^{4}$, and integrate out the  $\tau$-dependent modes, the end
result is a three-dimensional Landau-Ginzburg theory with effective
Hamiltonian
\beq
H[\Phi] = \int d\vecx\;[\frac{1}{2m}|\nabla\Phi|^{2} - \mu(T)|\Phi|^{2}
+\half\lambda |\Phi|^{4}].
\eeq
like that of (3.10).
At this level of approximation, in which Euclidean time has disappeared, the
distinction between relativistic and non-relativistic theories has
also disappeared (except for the details of $\mu (T)$).
But for a slight change of definition, repeating our analysis will
give a variance $(\Delta n)^{2}$ in winding number (along $\partial S$)
proportional
to the path length and to the radial field fluctuation.  In thermal
equilibrium there is, therefore, a true identity between the vortex
fluctuations of a relativistic scalar field of the type that could
occur in the early universe and those of superfluid $^{4}He$.  In the
Ginzburg regime, when a random phase distribution is relevant,
the vortex density will be large.

As a further example of scalar fields in thermal equilibrium
we observed earlier that if $\Pi_{2}(M)$ is non-trivial we can have
monopoles.  In a cosmological context they are an embarrassment \cite{TK},
since they would most likely dominate the energy of the universe if
produced at the GUT phase transition and allowed to survive.
As a final generalisation of our result (4.21) for winding number we
note that it is straightforward to repeat our analysis for the
production of global monoples.  Consider the simplest case of a vector
triplet of scalar fields in a broken global $O(3)$ theory.
The variance in the monopole charge $Q_{v}$ in a volume $v$
is, in the same approximation as the above, a product of three
two-point correlation functions.
The end result is \cite{AG} that, in thermal equilibrium at temperature $T$,
for coupling constant $\lambda$ and effective Higgs mass $m(T)$,
\beq
\expect{Q_{v}Q_{v}} = O((\lambda T/m(T))^{3})O((m(T)^{3}v)^{2/3})
\eeq
where $m^{3}v$ is the volume $v$ measured in terms of correlation
volumes, correlation length $\xi = m(T)^{-1}$.
In the Ginzburg regime this is understood as the field
taking random directions in field space in different correlation
volumes.

\subsection{Out-Of-Equilibrium Behaviour.}

Our discussion above has shown that, in thermal equilibrium close to the phase
transition, field
fluctuations are large enough to produce local topological charge
(winding number, monopole charge) at high density, as anticipated by
Kibble.  However, this is of little use unless the future evolution
of the system is able to freeze this charge in.  Let us return to
relativistic vortices.  The random distribution of field phases shown above
leads
to macroscopic vortex fluctuations, crossing the system volume $V$.  Numerical
simulations may take an initial configuration of vortices based on random
phases and
then, switching to zero (or low) temperature solve for their classical
evolution.
This freezing in of
defects goes beyond anything that we have said so far.  As with the
real scalar field of the previous section it is simplest to adopt a
quenched approximation in which the closed time-path action for $t >
0$ is $S$ of (4.1).  In the light of our previous comments one
of the more interesting choices would be to take the initial state
as one of thermal equilibrium in the Ginzburg regime with its large
fluctuations and watch to see if the vortices freeze.
However, a  simpler, and probably more relevant, choice is to move
instantaneously from a
symmetric to a spontaneously-broken theory, rather as in (3.16), but
with the differences outlined below.

Again we assume an initial symmetric state, centred at $\phi = 0$.  Our
first guess might be to take field configurations Boltzmann-distributed
as $P_{in}[\Phi] = N\exp\{-\beta_{in}H_{in}[\Phi]\}$, where
$H_{in}[\Phi]$ is derived from the free-field
action
\beq
S_{in}[\phi] = \int d^{4}x\;[\half|\partial\phi|^{2} - \half
m_{in}^{2}|\phi|^{2}]
\eeq
the $U(1)$ generalisation of (2.38).  However, when (4.22) is
decomposed into radial and angular fields as
\beq
S_{in}[\rho ,\chi] = \int d^{4}x\;[\half(\partial\rho)^{2}
+\half\rho^{2} (\partial\chi)^{2} -\half m_{in}^{2}\rho^{2}]
\eeq
the coupling between them (the second term) cannot be ignored since
$\expect{\rho} = 0$.  In the absence of any compelling reason
(initial thermal equilibrium in the symmetric phase
is hardly likely in the universe, although it would be appropriate
for superfluid $^{4}He$) we may as well make a choice of
initial condition that is solvable.  In this spirit, we take
\beq
S_{in}[\rho ,\chi] = \int d^{4}x\;[\half(\partial\rho)^{2}
+\half\sigma^{2} (\partial\chi)^{2} -\half m_{in}^{2}\rho^{2}]
\eeq
where $\sigma^{2} = m^{2}/\lambda$ determines the vacuum in the
subsequent symmetry-broken phase.

This $S_{in}$, which characterises the initial probabilities, is to be
contrasted to the action which determines the subsequent spinodal
evolution of the field.  A similar problem arises for times $t >
t_{in}$.  Our first thought,
the quadratic (in $\phi$) action with downturned potential
\beq
S[\rho ,\chi] = \int d^{4}x\;[\half(\partial\rho)^{2}
+\half\rho^{2} (\partial\chi)^{2} +\half m^{2}\rho^{2}]
\eeq
with $m_{H} = \sqrt{2} m$, involves interactions between the Higgs
and Goldstone fields after the radial/angular field decomposition.
In the spirit of the Gaussian approximation we replace it by
\beq
S[\rho ,\chi] = \int d^{4}x\;[\half(\partial\rho)^{2}
+\half\sigma^{2} (\partial\chi)^{2} +\half m^{2}\rho^{2}]
\eeq
in which the decoupling between $\rho$ and $\chi$ is enforced
by putting $\rho^{2} =\sigma^{2}$ in the second term of (4.29)
The identity of the massless $\chi$  action in (4.28) and (4.30)
guarantees that it remains in thermal equilibrium at all times.
Although $S$    of (4.30) is artificial,
it is realistic in that we would not expect  the  $\chi$  modes to be
dominantly exponential,
but oscillatory.
This is in contrast to the quenched behaviour of
the radial Higgs field $\rho$ as it falls off
the hill, with its exponentially growing long-wavelength fluctuations.
The Goldstone field, always being massless,
possesses the IR singularity for small $\vecp$,
\beq
G_{g}(\vec{p} ;t) \simeq \frac{T_{in}}{\vec{p}^{2}}
\eeq
With this $1/\vec{p}^{2}$ effectively cancelling the $p_{L}^{2}$ in
$\cal J$,  $\expect{\bar{N}_{S}(t)\bar{N}_{S}(t)}$
behaves as
\bea
\expect{\bar{N}_{S}(t)\bar{N}_{S}(t)} &\simeq&
{\cal J} \left(\frac{2}{2\pi\sigma^{2}}\right)mT_{in}
\int_{|\vec{p}|<m} \dbar\vec{p} G_{\rho}(\vec{p} ;t)
\\
&=&
O(L)\left(\frac{2}{2\pi\sigma^{2}}\right)mT_{in}\expect{\rho_{v}(t)\rho_{v}(t)}
\eea
where $L$ is the path length in units of $l$   which, now, we take to be $\xi$
of (3.25).

As before, the fluctuations in winding number vary
as the square root of the step number, a consequence of random phase
fluctuations, with scale set by the fluctuations of the Higgs field
at the cessation of domain growth.  However, with $\rho$ positive there is now
only one side of the hill
to roll down.  The non-vanishing of $<\rho_{v}>$ prevents the Higgs two-point
correlation function
being
identical to that of the scalar field in (3.21) and (3.24).
However, with $<\rho_{v}>$ itself showing exponential growth the effect may be
qualitatively
the same.  Details will be given elsewhere \cite{ER}.
How these domains then
aggregate to produce the true vacuum is beyond our calculations, as
is the subsequent evolution of the vortex network.  What is
interesting is the similarity between the equilibrium result (4.21)
and (4.33).
Whether the vortices are produced in equilibrium or strongly out of equilibrium
the
Kibble conjecture seems substantially correct.

Global monopoles can be treated in the same way.
If the Goldstone modes are allowed to stay in equilibrium the
fluctuations in monopole charge retain the equilibrium volume
dependence of (4.25), scaled by the Higgs fluctuations.
This is not necessarily the case for $^{4}He$. The form (4.23) for
the $^{4}He$ $\Phi$-field partition function relied
on thermal equilibrium for the particle system to be identical to a
field theory.  We do not know to what extent a non-equilibrium
system of point particles can be put into correspondence with
non-equilibrium field theory.  However, if it were possible to take
a real-time continuation of (4.23) as a basis for non-equilibrium
calculations there would be no difficulty in principle in adopting
the same approximations, although the difference between
non-relativistic and relativistic thermal field theories would now
have to be addressed.

\subsection{Local Vortices}

Our experience of continuous symmetries in particle physics has been
that all such symmetries have been made local by the presence of
gauge fields.  In that sense the global vortices that we have been
discussing are unnatural.

The local gauge extension of the $U(1)$ theory (4.1) has action
\beq
S[\phi,A] = \int d^{4}x\;[-\quarter F_{\mu\nu}F^{\mu\nu} +
|\partial_{\mu}\phi -ie A_{\mu}|^{2} + m^{2}|\phi|^{2}
- \quarter\lambda|\phi|^{4}].
\eeq
(changing factors of $\half$ for convenience).  This still permits
relativistic vortices, the simplest candidates for local cosmic
strings.
[Whereas the non-relativistic global strings are the vortices of
superfluids, the non-relativistic counterparts of local strings are
the vortices of superconductors].  For $e^{2}/\lambda\ll 1$ we have
a Type-II theory and the strings are approximately global, and our
previous results should apply.

The gauge-invariant expression (4.2) for winding number is still
valid, but it is more convenient to define it through the gauge
field $A_{\mu}$ as the line integral
\beq
N_{S}(t) = \frac{e}{2\pi}\int_{\partial
S}\underline{dl}.\underline{A}
\eeq
As before, field fluctuations will create local winding number.  In
the Gaussian approximation its variance is
\beq
\expect{N_{S}(t)N_{S}(t)} = \left(\frac{e}{2\pi}\right)^{2}
\int_{\partial S}dl'_{i}\int_{\partial
S}dl''_{j}\;\expect{A_{i}(t,\vecx')A_{j}(t,\vecx'')}
\eeq
where $\vecx',\vecx''$ denote the positions of the line increments
on $\partial S$.  In one sense this is significantly simpler than
its scalar counterpart (4.10) since we do not have to worry about
disconnected and connected parts.  The difficulty lies in the more
complicated correlation function.  There is no intrinsic problem in calculating
$\expect{N_{S}N_{S}}$ in thermal equilibrium.  At the same level of
approximation as before $\expect{N_{S}(t)N_{S}(t)} = O(e^{2}LT)$, up
to infrared logarithms.  This shows the usual $O(L)$ behaviour and,
in step lengths $\xi = m_{v}^{-1} = (e\sigma (T))^{-1}$,
$\expect{N_{S}(t)N_{S}(t)}$ shows large steps in winding number in
the Ginzburg regime.  Further, since the winding number is
essentially the magnetic flux a similar calculation can be performed
for estimating primordial magnetic fields \cite{KE}. However, as of this
moment the work is not complete, but we hope to give the results later.
This seems a good place to stop.

\section*{Acknowledgements.}

This talk is based, in large part, on collaborations with Mark Hindmarsh of
Sussex (equilibrium fluctuations), Tim Evans of I.C.
(non-equilibrium behaviour and gauge propagators) Luis Bettencourt of I.C.
(first-order transitions and gauge fluctuations) and Alistair Gill
(monopoles and $^{4}He)$. I thank them all.

\npagepub

\typeout{--- references ---}

\end{document}